\documentclass[%
 reprint,
superscriptaddress,
 amsmath,amssymb,
 aps,
]{revtex4-2}
\usepackage{xcolor}
\definecolor{apsblue}{RGB}{51,51,153}
\usepackage{natbib}
\usepackage[
    colorlinks=true,
    linkcolor=apsblue,   
    citecolor=apsblue,   
    urlcolor=apsblue     
]{hyperref}

\makeatletter
\def\NAT@spacechar{} % remove space after comma in [7,8]
\makeatother

\usepackage{physics,amsmath,amssymb,theorem,color,epsfig,epic,subfigure,hhline,graphicx}% Include figure files
\usepackage{dcolumn}% Align table columns on decimal point
\usepackage{bm}% bold math
\begin{document}

\preprint{APS/123-QED}

\title{Time-Multiplexed Distributed Quantum Sensing}
%Spatiotemporal Heisenberg Scaling via Time-Multiplexed Distributed Quantum Sensing\\}
%Time-Multiplezed Distributed Quantum Sensing with Spatiotemporal Heisenberg Scaling}% Force line breaks with \\

\author{Hanbom Yoo}%
\affiliation{%
 Department of Physics, Yonsei University, Seoul 03722, Republic of Korea}
 \author{Hyukgun Kwon}%
 \affiliation{Department of Physics and Astronomy, Sejong University, 209 Neungdong-ro Gwangjin-gu, Seoul 05006, Republic of Korea}
\author{Seongjin Hong}%
 \email{shong@yonsei.ac.kr}
\affiliation{%
 Department of Physics, Yonsei University, Seoul 03722, Republic of Korea}%
 \affiliation{%
 Department of Quantum Information, Yonsei University, Seoul 03722, Republic of Korea}

\date{\today}% It is always \today, today,

\begin{abstract}
Quantum metrology enables parameter estimation beyond classical limits by exploiting nonclassical resources such as squeezing and entanglement. In distributed quantum sensing, Heisenberg scaling has been extended from $1/N^2$ to $1/(NM)^2$ through entanglement across both particles and spatial modes, where $N$ denotes the photon number and $M$ the number of spatially distributed modes. However, the overall sensitivity has remained limited to linear scaling with the number of measurement repetitions $R$. Here, we show that exploiting entanglement across temporal modes via time-domain multiplexing enables a scaling advantage with respect to $R$. As a result, the sensitivity can asymptotically approach simultaneous Heisenberg scaling in photons, spatial modes, and repetitions, yielding an overall sensitivity approaching $\Delta^2 \phi \propto 1/(NMR)^2$. Using the Bogoliubov transformation formalism, we prove the optimality of the protocol within the class of Gaussian states and show that the scaling is realizable via homodyne detection and maximum-likelihood estimation. We further show that the advantage persists under optical loss and propose an experimentally feasible loop-based photonic sensing scheme. Our results open a route to incorporating time-multiplexing techniques into quantum metrology.
\end{abstract}

\maketitle

%\tableofcontents

\textit{Introduction---}
The goal of quantum metrology is to achieve parameter sensitivities beyond classical limits by exploiting quantum resources such as squeezing and entanglement~\cite{piran18,lawri19,polin20}. A central milestone in this field is the achievement of Heisenberg scaling (HS)~\cite{giova06}, where the estimation uncertainty scales as $1/N^2$ with the number of photons $N$, representing a quadratic improvement over the standard quantum limit (SQL), $1/N$~\cite{giova04}. This advantage has motivated extensive studies of quantum sensing protocols that maximize the extractable information under constrained resources~\cite{giova11,lupu22,hanam23,nair22,nair18,niels23,zhang25,andrade20}.

Distributed quantum sensing (DQS)~\cite{zhang21,pezze25} further extends this advantage by exploiting entanglement across spatially separated sensors. In particular, it has been shown that the sensitivity for estimating a global parameter—defined as a weighted combination of spatially distributed parameters—can scale as $1/(NM)^2$ by exploiting entanglement across both particles and spatial modes~\cite{gessn18}, where $M$ denotes the number of spatial modes. Numerous theoretical~\cite{gessn18,gessn20,oh20,oh22,matsu19,zhuang18,ge25,proct18,nehra24,zwierz10,kwon22,kwon19} and experimental~\cite{guo20,hong25,liu21,du25,hong21,kim25,kim24} studies have demonstrated that various quantum states of light can achieve this scaling, establishing DQS as a rapidly developing frontier of quantum metrology.

Despite these advances, the overall sensitivity typically exhibits only linear scaling with the number of measurement repetitions $R$, i.e., $\Delta^2 \phi \propto 1/R$. This limitation arises because most sensing protocols are optimized for single-shot measurements and implicitly assume that parameter estimation is performed independently across temporal modes, leading to only additive accumulation of Fisher information~\cite{paris09}. However, when multiple temporal modes can, in principle, be jointly accessed, such a measurement strategy imposes an unnecessary restriction and is generally not optimal.
Meanwhile, extensive theoretical and experimental efforts have been devoted to time-domain multiplexing that exploits entanglement in the temporal domain \cite{asava22,menic11}, including the generation of large-scale entangled states \cite{yoshi16,yoko13,lars19,asava19} and more sophisticated entangled structures~\cite{takeda19,takeda17,yonezu23,enomoto21}. While these developments provide a practical platform for generating and manipulating entanglement across temporal modes, their implications in quantum metrology have remained largely unexplored. In particular, improving the scaling with respect to the number of repetitions, such as $\Delta^2 \phi \propto 1/R^2$, has not yet been investigated.

\begin{figure*}[t]
\centering\includegraphics[width=\textwidth]{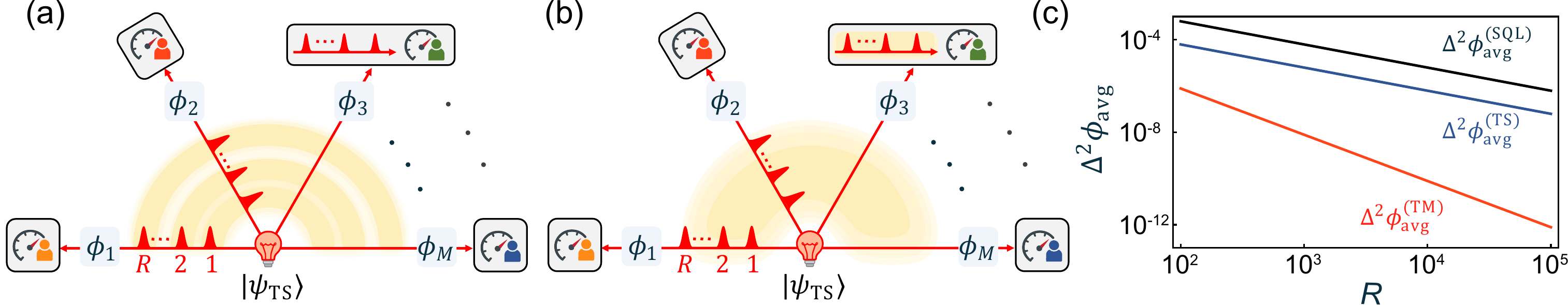}
    \caption{Distributed quantum sensing with (a) time-separable (TS) and (b) time-multiplexed (TM) protocols. 
The probe state $\hat{\rho}$ acquires spatial phase shifts $\{\phi_j\}$ and the average phase is inferred from correlated measurements. 
The shaded region indicates entanglement between spatiotemporal modes. 
(c) Variance of $\phi_{\mathrm{avg}}=\sum_{i=1}^{M}\phi_i/M$ versus the number of measurement repetitions $R$. 
The TM scheme exhibits improved scaling compared with the TS protocol and the standard quantum limit (SQL), with $\bar{n}=2$, $M=2$, and $\eta=1$ fixed.}
\label{fig1}
    \end{figure*}

In this Letter, we show that exploiting entanglement across temporal modes via time-domain multiplexing enables the scaling with respect to the number of measurement repetitions to asymptotically approach Heisenberg scaling. In the large-$R$ limit, the estimation error approaches $\Delta^2\phi \propto 1/R^2$, extending the previously established scaling $\Delta^2\phi \propto 1/(NM)^2$ toward an asymptotic simultaneous HS in the number of photons, spatial modes, and repetitions, yielding an overall sensitivity approaching $\Delta^2\phi \propto 1/(NMR)^2$. Using the Bogoliubov transformation formalism, we prove the optimality of the time-multiplexed protocol within the class of Gaussian probes by analyzing the quantum Cramér--Rao bound (QCRB) and the corresponding classical Cramér--Rao bound (CRB). We further show that homodyne detection constitutes an optimal measurement strategy that saturates the QCRB and that an estimator derived from homodyne outcomes realizes the advantage of time multiplexing. Moreover, the quantum enhancement persists under optical loss. Finally, we propose an experimentally feasible loop-based photonic sensing architecture that can be readily implemented on photonic platforms~\cite{takeda19,takeda17,yonezu23,enomoto21}.

\textit{Distributed phase sensing protocol---}
We now introduce the distributed phase-sensing protocol used to demonstrate HS with respect to the number of photons $N$, spatial modes $M$, and measurement repetitions $R$, as illustrated in Fig.~\ref{fig1}. Our goal is to estimate a global parameter 
$\phi=\sum_{j=1}^{M}\omega_j\phi_j$, defined as a weighted linear combination of spatially distributed phase shifts $\boldsymbol{\phi}=(\phi_1,\phi_2,\ldots,\phi_M)^{\mathsf T}$, where $\mathbf{w}=(\omega_1,\omega_2,\ldots,\omega_M)^{\mathsf T}$ is the weight vector.
We define the spatiotemporal probe mode as $\hat{a}_{t,j}\equiv\int_{-\infty}^{\infty} dt' f_t(t')\,\hat{a}_j(t')$,
where the mode function $f_t(t')$ is nonzero only within the time window $(t-1)\tau \le t' \le t\tau$~\cite{yoshi16}. The local phase shifts $\{\phi_j\}$ are encoded via the unitary
$\hat{U}_{\boldsymbol{\phi}}^{(t)}
=
\exp\!\left(
i\sum_{j=1}^{M}\phi_j\,\hat{a}_{t,j}^\dagger \hat{a}_{t,j}
\right)$
applied at each time step $t$. The resulting state after $R$ rounds is $\hat\rho_{\boldsymbol{\phi}}= \left(\bigotimes_{t=1}^{R}\hat{U}_{\boldsymbol{\phi}}^{(t)}\right)\hat \rho \left(\bigotimes_{t=1}^{R}\hat{U}_{\boldsymbol{\phi}}^{(t)}\right)^\dagger$.

Local measurements are then performed at each spatial node over $R$ rounds, and the collected outcomes are processed to estimate the global parameter. Using time-domain multiplexing, the probe state can be prepared as
$\hat{\rho}=\bigotimes_{r=1}^{R_r}\hat{\rho}^{(r)}$, 
where each $\hat{\rho}^{(r)}$ is defined the composite Hilbert space $\mathcal{H}_S^{\otimes R_t}$ and may exhibit entanglement across temporal modes~\cite{vanloock03,kim02}. Here $\mathcal{H}_S$ denotes the $M$-mode Fock space associated with the spatial modes. The total number of temporal modes per spatial channel is $R=R_rR_t$, which also corresponds to the total number of local measurements: each run produces $R_t$ time-correlated outcomes and is independently repeated $R_r$ times.

To highlight the role of temporal correlations, we compare the time-multiplexed (TM) strategy with $R_r=1$ and $R_t=R$ to the conventional time-separable (TS) scheme with $R_t=1$ and $R_r=R$, in which the probe state is a product state across temporal modes as commonly considered in previous work~\cite{guo20,liu21,oh20,oh22,matsu19,ge25,proct18}.
To quantitatively compare the TS and TM schemes, we analyze the ultimate precision limits imposed by quantum estimation theory. For multiparameter estimation of $\boldsymbol{\phi}$, the error matrix
$\Sigma_{ij}=\langle(\hat{\phi}_i-\phi_i)(\hat{\phi}_j-\phi_j)\rangle$
of any unbiased estimator is bounded by the CRB,
$\boldsymbol{\Sigma}\ge \mathbf{F}^{+}$,
where $\mathbf{F}$ is the Fisher information matrix and $+$ denotes the Moore–Penrose inverse~\cite{barata12}. Optimizing over all quantum measurements yields the QCRB, $\mathbf{F}^{+}\ge \mathbf{H}^{+}$, where $\mathbf{H}$ is the quantum Fisher information matrix (QFIM) defined via the symmetric logarithmic derivatives~\cite{paris09}.

For estimating a linear combination of parameters $\phi=\mathbf{w}^\top\boldsymbol{\phi}$, the estimation error obeys~\cite{li12}
\begin{equation}
\Delta^2 \hat\phi
= \langle (\hat{\phi}-\phi)^2 \rangle
\ge \mathbf{w}^\top \mathbf{F}^{+}\mathbf{w}
\ge \mathbf{w}^\top \mathbf{H}^{+}\mathbf{w},
\label{eq:bound}
\end{equation}
provided that the weight vector $\mathbf{w}$ lies within the support of both $\mathbf{H}$ and $\mathbf{F}$~\cite{kwon25,namkung24}. 

Due to the convexity of the QFIM~\cite{ge18}, it suffices to consider pure probe states. For a pure state $|\psi\rangle$, the generators of the parameters are
$\hat G_j \equiv i(\partial_{\phi_j}\hat U_{\boldsymbol{\phi}})\hat U_{\boldsymbol{\phi}}^\dagger
= \hat N_j$, where $\hat N_j=\sum_{t=1}^{R}\hat n_{t,j}$ and $\hat n_{t,j}=\hat a_{t,j}^\dagger\hat a_{t,j}$. The QCRB then reads~\cite{liu20}
\begin{equation}
\Delta^2\phi
\ge
\mathbf{w}^{\mathsf T}\mathbf{H}^{+}\mathbf{w}
=
\mathbf{w}^{\mathsf T}\!\left[4\,\mathrm{Cov}(\hat G_i,\hat G_j)_\psi\right]^{+}\!\mathbf{w},
\end{equation}
where $\mathrm{Cov}(\hat G_i,\hat G_j)_\psi$ denotes the covariance matrix of the generators evaluated on $|\psi\rangle$.

\textit{Quantum Fisher Information of Time-Multiplexed Gaussian Probes---}
To analyze the sensitivity achievable with time-multiplexed Gaussian probes, we introduce a composite mode index $\mu=(t,j)$, where $t\in\{1,\ldots,R\}$ labels temporal modes and $j\in\{1,\ldots,M\}$ labels spatial modes. Collecting the annihilation operators into the vector $\hat{\boldsymbol a}=(\hat a_{1,1},\ldots,\hat a_{1,M},\hat a_{2,1},\ldots,\hat a_{R,M})^{\mathsf T},$ we describe a total of $RM$ spatiotemporal modes. Any pure Gaussian state can be written as $|\psi\rangle=\hat D(\boldsymbol\alpha)\hat U_K \hat S(\boldsymbol r)|0\rangle$ with multimode displacement $\hat D(\boldsymbol\alpha)$, squeezing $\hat S(\boldsymbol r)$, and a passive interferometer $\hat U_K$ acting on the modes~\cite{weed12}. In the Heisenberg picture this corresponds to the Bogoliubov transformation~\cite{weed12,braun05} $\hat{\boldsymbol a}\mapsto U\hat{\boldsymbol a}+V\hat{\boldsymbol a}^\dagger+\boldsymbol\alpha$ where $U = WC$, $V = WS$ with $C = \mathrm{diag}(\cosh r_{1}, \ldots, \cosh r_{RM})$, $S = \mathrm{diag}(\sinh r_{1}, \ldots, \sinh r_{RM})$, and unitary matrix $W$. Using this transformation, the QFIM of a pure Gaussian state takes the form~\cite{liu20}
\begin{align}
H_{ij}
&=4\sum_{t,t'=1}^{R}\mathrm{Cov}(\hat n_{t,i},\hat n_{t',j})\\ \nonumber
&=4\sum_{t,t'=1}^{R}\mathrm{Cov}(\hat n_{i+(t-1)M},\hat n_{j+(t'-1)M}),
\end{align}
where the covariance is given by
\begin{align}
\mathrm{Cov}(\hat n_i,\hat n_j)
=&\, |(UV^{\mathsf T})_{ij}|^2+|(VV^\dagger)_{ij}|^2+\delta_{ij}(VV^\dagger)_{ji} \notag\\
&+\alpha_i\alpha_j^*(V^\dagger V)_{ij}
+\alpha_i^*\alpha_j(UU^\dagger)_{ij} \notag\\
&+\alpha_i^*\alpha_j^*(UV^{\mathsf T})_{ij}
+\alpha_i\alpha_j(UV^{\mathsf T})_{ij}^* ,
\end{align}
with further details provided in Supplemental Material~A.

To ensure experimental implementability we restrict the interferometer to $\hat U_K=\hat U_S\otimes\hat U_T$, corresponding to independent spatial and temporal interferometers. Under a fixed total photon-number constraint $\bar N=\sum_{l=1}^{RM}\langle\hat n_l\rangle$, the optimal time-multiplexed Gaussian probe is obtained for $R_r=1$ and $R_t=R$ [Fig.~1(b)], prepared by injecting a single-mode squeezed vacuum followed by a spatiotemporal passive interferometer, $|\psi\rangle=\hat U_K\bigl(\hat S(r)|0\rangle\otimes|0\rangle^{\otimes(RM-1)}\bigr)$.

For positive weights $w_j$ (negative weights are discussed in~\cite{oh22,ge25}), the resulting QFIM reads (see Supplemental Material~B)
\begin{equation}
H_{ij}=4\bar N(2\bar N+1)\,|(U_S)_{i1}|^2|(U_S)_{j1}|^2
+4\bar N\,\delta_{ij}|(U_S)_{i1}|^2 .
\end{equation}
Choosing the spatial interferometer such that
$|(U_S)_{j1}|^2=w_j/\sum_{k=1}^{M}w_k$,
while $\hat U_T$ remains arbitrary, yields the QCRB
\begin{align}
\Delta^2\phi
\ge \mathbf w^{\mathsf T}\mathbf H^{-1}\mathbf w
= \frac{(\sum_{k=1}^{M}w_k)^2}{8\,\bar N(\bar N+1)},
\end{align}
showing that the optimal time-multiplexed Gaussian probe achieves Heisenberg scaling under the total photon-number constraint.

\begin{figure}[t]
    \centering
    \includegraphics[width=0.95\columnwidth]{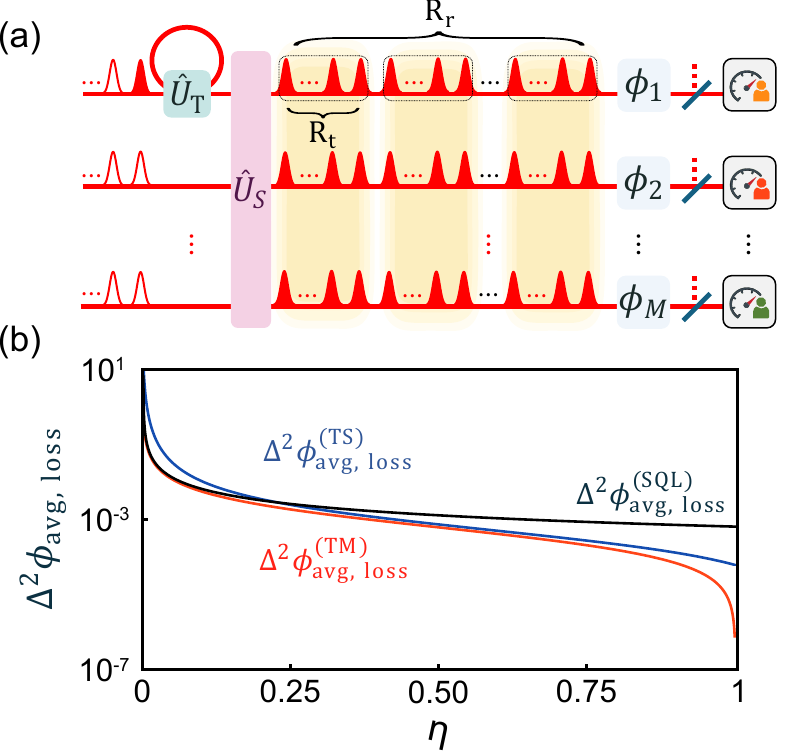}
    \caption{
(a) Time-multiplexed distributed sensing protocol. A squeezed probe is distributed across spatial modes via the interferometer $\hat U_S$ and across temporal modes via $\hat U_T$. The protocol is repeated $R_r$ times with $R_t$ temporal modes per run, giving a total number of measurements $R=R_rR_t$. 
(b) Sensitivity in the presence of optical loss $\eta$. The TM protocol retains a clear advantage over the TS scheme and the standard quantum limit (SQL) over a broad range of loss. Here, $\bar{n}=2$, $M=2$, and $R=10^{3}$, with $\eta=1$ corresponding to the lossless case.
    } 
    \label{fig2}
\end{figure}

\textit{Comparison of Time-Separable and Time-Multiplexed Schemes---}
To highlight the enhancement enabled by time multiplexing, we consider estimating the average phase $\phi_{\mathrm{avg}} = \tfrac{1}{M} \sum_{j=1}^{M} \phi_j$.
We first analyze the conventional TS scheme corresponding to $R_t=1$ and $R_r=R$ as shown in Fig.~\ref{fig1}(a), where temporal modes are independent and only a spatial interferometer is employed. The optimal Gaussian probe state is~\cite{oh20}
\begin{equation}
\ket{\psi_{\mathrm{TS}}}
=
\bigotimes_{t=1}^{R}
\left[
\hat U_S
\bigl(
\ket{r_{\mathrm{TS}}}_{t,1}
\otimes
\ket{0}_{t,2}
\otimes
\cdots
\otimes
\ket{0}_{t,M}
\bigr)
\right],
\end{equation}
where $\ket{r_{\mathrm{TS}}}=\hat S(r_{\mathrm{TS}})\ket0$ is a single-mode squeezed vacuum injected into one spatial mode at each time step. Choosing $r_{\mathrm{TS}}=\sinh^{-1}\!\sqrt{M\bar n}$ and $|(U_S)_{j1}|^2=1/M$ fixes the mean photon number per spatiotemporal mode to $\bar n$. Substituting $\bar N=M\bar n$ into Eq.~(7) and accounting for $R$ independent repetitions yields
\begin{equation}
\Delta^{2}\phi_{\mathrm{avg}}^{(\mathrm{TS})}
\ge
\frac{1}{8\left(RM^{2}\bar n^{2}+RM\bar n\right)} .
\end{equation}
This bound exhibits Heisenberg scaling (HS) with respect to $M$ and $\bar n$, but only linear scaling with the number of repetitions $R$.

We next consider the optimal TM scheme, where entanglement across temporal modes is allowed. A single squeezed vacuum is injected into the mode $(t,j)=(1,1)$ and distributed across temporal and spatial modes via passive interferometers,
\begin{equation}
\ket{\psi_{\mathrm{TM}}}
=
\hat U_S \hat U_T
\bigl(
\ket{r_{\mathrm{TM}}}_{1,1}
\otimes
\ket0^{\otimes(RM-1)}
\bigr).
\end{equation}
Choosing $r_{\mathrm{TM}}=\sinh^{-1}\!\sqrt{RM\bar n}$ and interferometers satisfying
$|(U_T)_{t1}|^2=1/R$ and $|(U_S)_{j1}|^2=1/M$
ensures the same local photon-number constraint $\bar n$. Equation~(7) then gives
\begin{equation}
\Delta^2\phi_{\mathrm{avg}}^{(\mathrm{TM})}
\ge
\frac{1}{8\left(R^{2}M^{2}\bar n^{2}+RM\bar n\right)} .
\end{equation}
The TM scheme therefore achieves HS simultaneously in $\bar n$, $M$, and $R$, enabled by entanglement across temporal modes as shwon in Fig.~\ref{fig1}(b). Figure~\ref{fig1}(c) compares the sensitivities with the standard quantum limit (SQL) obtained using coherent states $\ket{\sqrt{\bar n}}^{\otimes RM}$, $
\Delta^{2}\phi_\text{avg}^\text{(SQL)} \ge 1/(4\eta RM\bar n)$. The TM scheme surpasses both the TS protocol and the SQL as $R$ increases, approaching HS with respect to the number of measurement repetitions.

\textit{Optimal Measurement and Estimation Error---}
We next analyze the measurement strategy required to saturate the QCRB. 
Under lossless conditions, homodyne detection on each mode with appropriately chosen local-oscillator phases asymptotically saturates the QCRB with respect to $R$ in the TS scheme~\cite{oh20}. However, in the time-multiplexed setting the homodyne outcomes are temporally correlated, and therefore the standard analysis for independent and identically distributed outcomes does not directly apply. We first consider the optimal-point estimation problem where $\phi_j=\phi_{\text{opt}}=\sin^{-1}(1/(2R_tM\bar{n}+1))/2$ is fixed. 
In this case, a locally unbiased estimator constructed from homodyne measurements saturates the QCRB using a single time-multiplexed probe, \textit{i.e.,} $R_r=1$ and $R_t=R$ (see Supplemental Material~C).

We next consider a more realistic local-estimation regime where the phases deviate slightly from $\phi_{\text{opt}}$. 
Specifically, we assume $\phi_{\mathrm{avg}}-\phi_{\mathrm{opt}}=C/R_t M \bar n$ and $
\sum_{j=1}^{M}(\phi_j-\phi_{\mathrm{avg}})^2/M=D/R_t M \bar n$,  with constants $C$ and $D$. 
In this regime, it becomes advantageous to repeat the TM protocol independently over $R_r$ runs, each containing $R_t$ temporal modes, so that the total number of measurements is $R=R_rR_t$.
Parameterizing \(R_r=R^{1-s}\) and \(R_t=R^{s}\) with \(s\in(0,1)\), the variance of the maximum-likelihood estimator (MLE) constructed from the homodyne outcomes is, for \(R\gg 1\), given by
\begin{equation}
\Delta^2\hat{\phi}_{\mathrm{avg, HD}}
=
\frac{f(C,D)}{8R^{1+s}M^2\bar n^2},
\end{equation}
where $f(C,D)$ is a constant factor (see Supplemental Material~C). Since $s$ can be chosen arbitrarily close to $1$, the estimator exhibits asymptotic scaling approaching the HS with respect to $R$. 
Moreover, for sufficiently large $R$, the TS scheme can be outperformed even when only a small fraction of the total measurements is allocated to independent repetitions. These results show that the quantum advantage predicted by the time-multiplexed protocol can be realized in practice using homodyne detection and the MLE.

It is noteworthy that the proposed TM metrology scheme is compatible with current photonic technology. 
Sequential temporal modes can be generated either from continuous-wave squeezed light with post-processed homodyne detection~\cite{yoshi16,kawasaki25,lars19,asava19,liu25} or from pulsed squeezed sources~\cite{gerrits11}. 
Temporal interferometers required for the protocol can be implemented using loop-based architectures that realize beam-splitter interactions between successive temporal modes~\cite{takeda19,takeda17,yonezu23,enomoto21,carosini24,motes14}, while the spatial interferometer $\hat{U}_S$ can be implemented using standard multiport beam-splitter networks in bulk, fiber, or integrated photonic platforms~\cite{carine20,clements16,bell21}. 
Further implementation details are provided in Supplemental Material~D.

\textit{Effects of Optical Loss---}
To assess realistic sensing conditions, we consider optical loss occurring independently in each spatial mode, as illustrated in Fig.~\ref{fig2}(a). Such loss may arise from photon leakage into the environment or imperfect detector efficiency~\cite{oh19}. The corresponding loss channel can be modeled by a beam-splitter interaction with environmental vacuum modes,
$\mathcal{L}_{\eta}(\hat{\rho}) =
\mathrm{Tr}_{\mathrm{E}}
\!\left[
\left(\bigotimes_{l=1}^{RM}\hat U_l(\eta)\right)
\left(\hat{\rho}\otimes|0\rangle\!\langle0|_{\mathrm E}^{\otimes RM}\right)
\left(\bigotimes_{l=1}^{RM}\hat U_l(\eta)\right)^\dagger
\right],$
where $\eta$ denotes the transmission and $\hat U_l(\eta)$ is the beam-splitter unitary~\cite{ferraro05}.
Under loss the Gaussian probe becomes mixed, and the QFIM for pure states is no longer applicable. 
For a mixed state $\mathcal{L}_{\eta}(\hat{\rho}_{\boldsymbol\phi})=\sum_m \lambda_m|\psi_m\rangle\!\langle\psi_m|$, the QFIM is given by~\cite{liu20}
\begin{equation}
H_{ij}
=
\sum_{m,n}
\frac{
2\,\mathrm{Re}\!\left(
\langle\psi_m|\partial_{\phi_i}\rho_{\boldsymbol\phi}|\psi_n\rangle
\langle\psi_n|\partial_{\phi_j}\rho_{\boldsymbol\phi}|\psi_m\rangle
\right)
}{
\lambda_m+\lambda_n
},
\end{equation}
where the sum excludes $\lambda_m+\lambda_n=0$.

For the time-multiplexed Gaussian probe considered here, the lossy state can be mapped to a single-mode squeezed thermal state followed by the spatiotemporal interferometer $\hat U_K$ (see Supplemental Material~E). This allows the QFIM to be evaluated analytically, yielding
\begin{equation}
\begin{aligned}
H_{ij} =
&\frac{4\eta\bar N(2\eta^2\bar N+2\eta-1)}
{1+2\eta(1-\eta)\bar N}
|(U_S)_{i1}|^2|(U_S)_{j1}|^2 \\
&+4\eta\bar N\,\delta_{ij}|(U_S)_{i1}|^2 .
\end{aligned}
\end{equation}

The resulting QCRB becomes
\begin{align}
\Delta^2\phi_{\text{loss}}
\ge
\frac{(1+2\eta(1-\eta)\bar N)\left(\sum_{k=1}^{M}w_k\right)^2}
{8\eta^2\bar N(\bar N+1)} .
\end{align}
For estimating the average phase $\phi_{\mathrm{avg}}$, the TS and TM schemes yield
\begin{align}
\Delta^{2}\phi_{\text{avg,loss}}^{(\mathrm{TS})}
&\ge
\frac{1+2\eta(1-\eta)M\bar n}
{8\eta^2 RM\bar n(M\bar n+1)}, \\
\Delta^{2}\phi_{\text{avg,loss}}^{(\mathrm{TM})}
&\ge
\frac{1+2\eta(1-\eta)RM\bar n}
{8\eta^2 RM\bar n(RM\bar n+1)} .
\end{align}

As expected, Heisenberg scaling is no longer preserved in the presence of optical loss. Nevertheless, the TM scheme retains a clear advantage over the TS protocol across the full range of $\eta$, as shown in Fig.~\ref{fig2}(c). In particular, the TM scheme surpasses the standard quantum limit (SQL) in the moderate-loss regime $\eta\ge1/2$, and also for $\eta<1/2$ when $RM\bar n\ge (1-2\eta)/(2\eta^2)$.

%Finally, in the presence of loss the optimal Gaussian measurement consists of a passive interferometer followed by general-dyne detection~\cite{serafini23}, which asymptotically saturates the QCRB of Eq.~(11) in the large-photon-number limit~\cite{oh20,oh19}.

\textit{Discussion---}
In summary, we have investigated the fundamental sensitivity limits of time-multiplexed quantum metrology with Gaussian probes. 
By analyzing the quantum Fisher information, we showed that temporal-mode entanglement enables a scaling advantage with respect to the number of measurement repetitions and allows the sensitivity to asymptotically approach Heisenberg scaling in the combined resources of photons, spatial modes, and repetitions. 
We further demonstrated that this advantage is attainable using homodyne detection with maximum-likelihood estimation and remains robust in the presence of optical loss.

The proposed TM protocol is particularly attractive for sensing scenarios where optical power must be limited, such as measurements constrained by quantum backaction~\cite{clerk10} or photodamage in biological samples~\cite{casac21}. 
More broadly, our results suggest that time-domain multiplexing provides a powerful new resource for quantum metrology and may enable enhanced performance in a wide range of sensing tasks, including distributed displacement estimation, optical loss or gain estimation, and quantum-channel characterization.

\begin{acknowledgments}
This work was partly supported by National Research Foundation of Korea (NRF) grant
funded by the Korea government (MSIT) (RS-2024-00352325, RS-2023-00283146), Global - Learning \& Academic research institution for Master’s·PhD students, and Postdocs(LAMP) Program of the National Research Foundation of Korea(NRF) grant funded by the Ministry of Education (RS-2024-00442483), and Institute 
for Information \& communications Technology Planning \& Evaluation (IITP) grant funded by the Korea government
(MSIT) (RS-2022-II221029, RS-2025-25464481).
\end{acknowledgments}

\nocite{*}
\bibliography{apssamp}% Produces the bibliography via BibTeX.

\end{document}